\documentclass[english,aps,superscriptaddress,amsmath,amssymb,longbibliography,onecolumn,floatfix,10pt]{revtex4-1}  
\usepackage{amsmath}
\usepackage{graphicx}  
\usepackage{dcolumn}   
\usepackage{bm}        
\usepackage{amssymb}   
\usepackage[normalem]{ulem}

\usepackage{tikz}
\usetikzlibrary{arrows, arrows.meta}
\usepackage{hyperref}
\hypersetup{
    colorlinks=true,
    linkcolor=blue,
    filecolor=blue,      
    urlcolor=black,
    citecolor=blue,
        }

\begin{document}

\title{A Reduced Model for a Phoretic Swimmer}

\author{Alexander Farutin}
\email{chaouqi.misbah@univ-grenoble-alpes.fr}
\affiliation{Univ. Grenoble Alpes, CNRS, LIPhy, F-38000 Grenoble, France}
\author{Suhail M. Rizvi}
\affiliation{Univ. Grenoble Alpes, CNRS, LIPhy, F-38000 Grenoble, France}
\affiliation{Current address: Department of Biomedical Engineering, Indian Institute of Technology Hyderabad, Sangareddy, Telangana 502285, India} 
\author{Wei-Fan Hu}
\affiliation{ Department of Mathematics, National Central University, 300 Zhongda Road, Taoyuan 320, Taiwan}
\author{Te-Sheng Lin}
\affiliation{Department of Applied Mathematics, National Chiao Tung University,
1001 Ta Hsueh Road, Hsinchu 300, Taiwan
} 
\author{Salima Rafai}
\affiliation{Univ. Grenoble Alpes, CNRS, LIPhy, F-38000 Grenoble, France}
\author{Chaouqi Misbah}
\affiliation{Univ. Grenoble Alpes, CNRS, LIPhy, F-38000 Grenoble, France}





\begin{abstract}
We consider a 2D model of an autophoretic particle in which the particle has a circular shape and emits/absorbs a solute that diffuses and is advected by the suspending fluid. Beyond a certain emission/absorption rate (characterized by a dimensionless  P\'eclet number, $Pe$) the particle is known to undergo a bifurcation from a non motile to a motile state, with different trajectories, going from a straight to circular and to a chaotic motion by progressively increasing  $Pe$. From the full model involving solute diffusion and advection, we derive a reduced closed model which involves only two time-dependent amplitudes $C_1(t)$ and $C_2(t)$ corresponding to the first two Fourier modes of the solute concentration field. This model consists of two coupled nonlinear ordinary differential equations for $C_1$ and $C_2$ and presents several great advantages:(i) the straight and circular motions can be handled fully analytically, (ii) complex motions such as chaos can be analyzed numerically very efficiently in comparison to the numerically expensive full model involving partial differential equations, (iii) the reduced model has a universal form dictated only by symmetries, (iv) the model can be extended to higher Fourier modes. The derivation method is exemplified for a 2D model, for simplicity, but can easily be extended to 3D, not only for the presently selected  phoretic model, but also for any model in which chemical activity triggers locomotion. A typical  example can be found, for example, in the field of cell motility involving acto-myosin kinetics. This strategy offers an interesting way to cope with swimmers on the basis of ordinary differential equations, allowing for analytical tractability and efficient numerical treatment.

\end{abstract}
\maketitle

\section{Introduction}
Phoretic particles (rigid particles  or drops) have been attracting recently an increased interest theoretically, numerically and experimentally \cite{izri2014self,MLB13,michelin_lauga_2014,schmitt2013swimming,Jin2017,morozov2019nonlinear,Morozov2019JFM,Hu2019,Morozov_soft,Chen2021,Hokmabad2021,Izzet2020}. In its simplest version, the model consists of a particle that emits or absorbs (with an emission/absorption rate $A$) a solute which diffuses, with bulk diffusion constant $D$,  and is advected by the suspending fluid.   The interaction between the solute and the particle can be shown \cite{michelin_lauga_2014} to result in a tangential flow along the particle,  in the form of (in the frame moving with particle) ${\bf u} \sim M \nabla _s c$, where $\bf u$ is the velocity, $c$ the solute concentration, $\nabla_s$ is the gradient along the particle surface, and $M$ is a mobility factor involving fluid viscosity and the interaction potential between the solute and  the particle. This problem can be characterized by a P\'eclet number $Pe=AM a/D^2$ where $a$ is the particle radius. The solute $c({\bf r},t)$ (${\bf r }$  is a position in space) and velocity ${\bf u} ({\bf r}, t)$ obey the advection-diffusion and the Stokes equations. It has been shown analytically, from a linear stability analysis, both in 2D \cite{Hu2019} and 3D \cite{MLB13} that if $Pe<Pe_1$ (where $Pe_1$ is a critical number), the particle undergoes a bifurcation from a non-motile state (swimming velocity $u_0=0$) into a motile state ($u_0\ne 0$). A linear stability only informs us on the instability onset from one state to another, but is not sufficient to describe how the velocity behaves with $Pe$, where a nonlinear analysis is needed. Numerical simulations showed\cite{MLB13,Hu2019} that the behavior of $u_0$ is well represented by ${ u}_0  \sim \pm (Pe-Pe_1)^{1/2}$ in the vicinity of the bifurcation point; the solution $u_0=0$ always exists, and is stable for $Pe<Pe_1$ and becomes  unstable for $Pe>Pe_1$, in favor of two stable branches of solutions ${ u}_0  \sim\pm (Pe-Pe_1)^{1/2}$. This is a classical pitchfork bifurcation. Numerical simulations in 2D \cite{Hu2019} also showed that by increasing $Pe$ further the straight moving solution becomes unstable in favor of  various states (meandering, circular, and chaotic solutions). Simulations in 3D  under axisymmetric constraint (imposing to the particle to move along a line)  also reported \cite{morozov2019nonlinear}on chaotic solutions (the particle goes back and forth in a chaotic manner). Relaxing the axisymmetric constraint in 3D also revealed, among other motion, meandering, and chaotic motion in the form of persistent random walk \cite{Hu3D}.  The transition from a non motile to a motile state, as well as irregular motion (apparently chaotic) has also been reported experimentally\cite{Hokmabad2021,Jin2017,izri2014self,Izzet2020}, and in some numerical simulations \cite{Chen2021}. The transition from a non motile to a motile state also takes place in the presence of a Marangoni stress \cite{izri2014self,morozov2019nonlinear}. In a completely different context, that of cell motility driven by acto-myosin rich dynamics (transition from a motile to a non motile state, Hopf bifurcation...) have also been reported when the activity of cell (represented by myosin contractility) exceeds a certain critical value \cite{hawkins2011spontaneous,Voituriez2016,Farutin2019}.

The above  non trivial behaviors have been obtained in most cases by numerical simulations of model equations involving diffusion, and advection with a moving boundary. It is thus highly desirable to see wether or not these features can be translated into a more universal language. One way is to seek for a reductive perturbative scheme that allows to extract from the full model prototypical simplified equations  which are universal (in the sense that their form does not depend on the considered swimmer model). 
This is the main goal of this paper. Our main objective here is, by starting from the above described phoretic model, to extract, by following a nonlinear perturbative scheme, two coupled (weakly nonlinear) equations for two complex amplitudes of the first two Fourier modes, which depend only on time, while the spatial dependence can be handled analytically. These equations will enable us to find, besides the first (primary) bifurcation (transition from a motile to a nonmotile state) a secondary bifurcation in the form of a circular trajectory. We will also exemplify its power in capturing irregular motion. We will exemplify the method for a 2D model (for simplicity), but the technique should work perfectly  well in 3D, not only for the phoretic model, but also for any model where locomotion is assisted by chemical activity.

\section{Problem formulation}
\label{fullmodel}
Our starting point is a simple phoretic model in 2D.
We first recall the problem formulation\cite{MLB13,michelin_lauga_2014}.
We consider a concentration field $c(\boldsymbol r,t)$ subject to an advection-diffusion equation
\begin{equation}
\label{advection-diffusion}
\partial_t c(\boldsymbol r,t)=D\nabla^2 c(\boldsymbol r,t)-\boldsymbol\nabla\boldsymbol \cdot [c(\boldsymbol r,t)\boldsymbol u(\boldsymbol r,t)],
\end{equation}
where the advection velocity field $\boldsymbol u(\boldsymbol r,t)$ in the co-moving frame is obtained by solving the Stokes equations in the fluid outside the particle.
Here $t$ is the time, $\boldsymbol r$ is the 2D position vector, measured relative to the particle center, and $D$ is the diffusion coefficient.
We use polar coordinates $(r,\phi)$, centered at the particle position in the following derivation.
The boundary conditions for the concentration field read
\begin{equation}
\label{bcc}
\partial_rc(\boldsymbol r,t)|_{r=a}=-A/D,\,\,\,c(\boldsymbol r,t)|_{r=R}=0,
\end{equation}
where $a$ is the particle size and $R$ is the system size.
The flow velocity satisfies the following boundary conditions:
\begin{equation}
\label{flow}
\boldsymbol u(\boldsymbol r,t)|_{r=\infty}=-\boldsymbol u_0(t),\,\,\,\boldsymbol u(\boldsymbol r,t)|_{r=a}={M\over a} \boldsymbol \nabla^s c(\boldsymbol r,t)|_{r=a},
\end{equation}
where $\boldsymbol u_0(t)$ is the swimming speed of the particle, $\boldsymbol\nabla^s$ is the surface gradient operator, and $M$ is the particle mobility.
The system is closed by requiring the net force acting on the particle to be zero, which allows one to compute the swimming velocity $u_0(t)$.
It is conventional to express the fluid velocity in terms of the stream function $\psi$
\begin{equation}
\label{stream}
u_r(\boldsymbol r,t)=\frac{\partial_\phi\psi(\boldsymbol r,t)}{r},\,\,\,u_\phi(\boldsymbol r,t)=-\partial_r \psi(\boldsymbol r,t).
\end{equation}
The stream function is related to the concentration distribution along the particle boundary by the boundary condition (\ref{flow}), as explained below.

Hereafter, we use a non-dimensional form of the problem, in which the size of the particle $a$, the diffusion coefficient $D$, and the release rate $A$ are all set to 1.
The remaining two non-dimensional numbers are $Pe=AMa/D^2$ and system size $R/a$. We will keep notation $R$ for this dimensionless quantity.

\section{Expansion in Fourier harmonics}

Following the previous works \cite{MLB13,Hu2019}, the concentration field and the stream function are expanded in Fourier harmonics of the polar angle $\phi$.
\begin{equation}
\label{Fourier}
c(\boldsymbol r,t)=\sum\limits_{l=-\infty}^\infty c_l(r,t)e^{il\phi},\,\,\,\psi(\boldsymbol r,t)=\sum\limits_{l=-\infty}^\infty \psi_l(r,t)e^{il\phi},
\end{equation}
where we have $c_l(r)=c_{-l}(r)^*$ and $\psi_l(r)=\psi_{-l}(r)^*$ as requirement for $c$ and $\psi$ to be real.
The Stokes equations and the boundary condition (\ref{flow}) allow one to express the amplitudes $\psi_l$ as
\begin{equation}
\label{psil}
\psi_l(r)=\frac{ilPe(1-r^2)}{2r^{|l|}}c_l(1).
\end{equation}
Exploiting the force-free condition on the particle, the swimming velocity is obtained as \cite{Hu2019} $v_x=-Pe\Re c_1(1)$, $v_y=Pe\Im c_1(1)$.

The advection-diffusion equation (\ref{advection-diffusion}) is expanded in Fourier harmonics as
\begin{equation}
\label{cldot}
\partial_t c_l(r)=\hat D_l c_l(r)-[\boldsymbol\nabla\boldsymbol\cdot(c(\boldsymbol r)\boldsymbol u(\boldsymbol r))]_l,
\end{equation}
where $\hat D_l$ is the diffusion operator applied to the $l$-th Fourier harmonic:
\begin{equation}
\label{Dl}
\hat D_l=\frac{1}{r}\partial_r(r\partial_r)-\frac{l^2}{r^2}
\end{equation}
and $[\boldsymbol\nabla\boldsymbol\cdot(c(\boldsymbol r)\boldsymbol u(\boldsymbol r))]_l$ is the l-th Fourier harmonic of the advection term.

\section{Stationary state and linear approximation}

The linear stability of the isotropic solution was already analyzed in previous works in 3D \cite{MLB13} and in 2D \cite{Hu2019}.
Nevertheless, we present it briefly as the first step of the solution of the non-linear problem.
The isotropic solution corresponds to 
\begin{equation}
\label{c0}
c_0(r)=-\ln\frac{r}{R}
\end{equation}
Substituting eq. (\ref{Fourier}) with $c_0(r)$ given by eq. (\ref{c0}) into eq. (\ref{cldot}) and neglecting terms non-linear in $c_l$, yields the following evolution equation for $c_l$
\begin{equation}
\label{linear}
\partial_t c_l(r)=\hat L_l(Pe) c_l(r)=\hat D_l c_l(r)+Pe u_l(r)c_l(1),
\end{equation}
where $\hat L_l$ is the linear stability operator for the $l$-th harmonic.
Here $Pe u_l e^{il\phi}$ is the advection of the unperturbed solution (\ref{c0}) with velocity field driven by the surface concentration $c_l(1)e^{il\phi}$:
\begin{equation}
\label{ul}
u_l(r)=\frac{l^2(r^2-1)}{2r^{l+2}}.
\end{equation}

The linear stability of the isotropic solution is governed by the eigenvalues $\lambda_{l,k}$ of the operators $\hat L_l$, defined as
\begin{equation}
\label{proper}
\hat L_l f_{l,k}(r)=\lambda_{l,k}f_{l,k}(r),
\end{equation}
where $f_{l,k}$ are the corresponding proper functions. The subscript $l$ is associated with the $\phi$ variable and $k$ with the $r$ one.
We index the eigenvalues for given $l$ in descending order with respect to their real part, starting with $\lambda_{l,0}\equiv\lambda_l$, which corresponds to the most unstable mode.
It is essential for our analysis to assume the eigenvalue spectrum to be discrete, which is the case for a finite domain.
In general, neither the eigenvalues nor the proper functions of $\hat L_l$ have an elementary expression.
However, it is known that for each $l\ge 1$, there is exactly one critical Peclet number $Pe_l$ such that the operator $\hat L_l(Pe_l)$ has an eigenvalue equal to zero\cite{michelin_lauga_2014,Hu2019}.
Both $Pe_l$ (i.e. $Pe_1$ and $Pe_2$) as a function of $R$ and the proper function of $\hat L_l(Pe_l)$ corresponding to the zero eigenvalue have an explicit expression in elementary functions, given below.

 For low enough $Pe$, the eigenvalues of $\hat L_l$ are close to those of the diffusion operator $\hat D_l$ and are thus all negative.
The isotropic solution is stable in this case (growth rate of all modes is negative for all $l$).
As $Pe$ increases, one of the proper values for given $l$ ($\lambda_{l,0}\equiv\lambda_l$) becomes equal to zero at $Pe=Pe_l$ and positive for $Pe>Pe_l$.
The corresponding proper function $f_{l,0}(r)\equiv f_l(r)$ defines the perturbation mode that becomes unstable at $Pe=Pe_l$.
The linear stability analysis allows us to determine the angular dependence (given by $l$ corresponding to the lowest $Pe_l$) and the radial dependence (given by $f_l(r)$) of the solution in the anisotropic phase.
Further analysis is performed by higher-order perturbation expansion in order to compute the swimming speed and the angular velocity of the particle in the anisotropic phase, as explained below.

Here we exclude the possibility of transition to an anisotropic concentration distribution by a Hopf bifurcation.
Since the operators $\hat L_l$ are not self-adjoint, their proper values can be complex.
It is therefore possible for the isotropic solution to become unstable due to one of the complex eigenvalues seeing its real part to become positive.
We exclude this possibility based on the results of full numerical simulations, shown below.

\section{Weakly non-linear expansion}

This section presents the derivation of the weakly non-linear equations governing the particle dynamics.
We  consider the situation where both the first and the second harmonic are close to the instability, or, equivalently, $Pe$ is simultaneously close to $Pe_1$ and $Pe_2$. This allows one to keep only thee first two leading harmonics. But It will appear that extension to other higher harmonics is straightforward, albeit we restrict ourselves here to keeping only the two first harmonics.
This is made possible by choosing an appropriate size of the system $R$.
We find that $Pe_1$ becomes equal to $Pe_2$ at $R=R_c$, where $R_c=3.17493$.
We thus have to assume that $R=R_c+O(\epsilon)$, but also $R>R_c$, the second condition guaranteeing $Pe_1<Pe_2$.
Here $\epsilon$ is the small parameter with respect to which the perturbation expansion is made.
With these assumptions, $c=c_0(r)$ is the only possible solution for $Pe<Pe_1$.
This solution becomes unstable at $Pe=Pe_1$ and a straight motion is expected to emerge at this point.
At some value of $Pe$ between $Pe_1$ and $Pe_2$, the straight motion is expected to become unstable in favor of a circular motion\cite{Hu2019}.
We have been able to compute the $O(\epsilon^2)$ terms analytically for an arbitrary value of $R$ and also all $O(\epsilon^3)$ terms for a given numerical value of $R$.
This is sufficient to reproduce quantitatively the straight and circular motions, as shown below. Our reduced model accounts also for irregular motions, reported in\cite{Hu2019}.

\subsection{General strategy}

The goal of this derivation is to reduce the full dynamics of the autophoretic particle to a simplified system of ordinary differential equations for two complex variables, $C_1$ and $C_2$, where $C_1$ is the amplitude of the first harmonic and $C_2$ is the amplitude of the second harmonic in the concentration field.
The main challenge is that in general, each Fourier harmonic $c_l$ of the concentration field is a function of the distance from the center of the particle $r$.
There is thus no straightforward way to represent the whole function $c_l$ by a single scalar variable.
We overcome this problem by decomposing the functions $c_1(r)$ and $c_2(r)$ as
\begin{equation}
\label{split}
c_l(r,t)=C_l(t)f_l(r)+\delta c_l(r,t)\,\,\,l\in\{1,2\},
\end{equation}
where $C_l(t)$ is the complex amplitude, $f_l(r)$ is the proper function such that $\hat L_l(Pe_l) f_l(r)=0$, and $\delta c_l(r,t)$ is a projection of the function $c_l(r,t)$ on the space of all other proper functions of the operator $\hat L_l(Pe_l)$:
\begin{equation}
\label{deltac}
\delta c_l(r,t)=\sum\limits_{k>0}C_{l,k}(t)f_{l,k}(r).
\end{equation}
The decomposition (\ref{split}) is made using the proper functions of $\hat L_l(Pe_l)$ even for $Pe\ne Pe_l$.
This is to capitalize on the explicit expression for $Pe_l$ and $f_l$ for $\hat L_l(Pe_l)$.
The difference between the linear stability operator $\hat L_l(Pe)$ and $\hat L_l(Pe_l)$ is of order $O(\epsilon)$ and is duly accounted for as a higher-order correction during the derivation.

The eigenvalues corresponding to the functions $f_{l,k}$ with $k>0$ all have negative real parts.
Therefore, the relaxation time scale of the amplitudes $C_{l,k}(t)$ is defined by the absolute values of $\lambda_{l,k}$.
Since the set of $\lambda_{l,k}$ is discrete and we have assumed $\lambda_{l,k}$ to be ordered decreasingly, the longest relaxation time scale is defined by $|\lambda_{l,1}|$.
The relaxation time of the $f_l(r)$ mode scales as $1/|Pe-Pe_l|$ for $Pe$ close to $Pe_l$, which determines the main time scale of the dynamics close to the critical point.
We thus have been able to reexpress the functions $\delta c_l(r,t)$ as a non-linear function of $C_1$ and $C_2$, with $r$-dependent coefficients.
This is done by adiabatic elimination, as explained below.
Since the eigenvalues and the proper functions of the operator $\hat L_l(Pe_l)$ do not have an elementary expression, the following procedure relies on the adiabatic elimination of the amplitude $\delta c_l(r,t)$ as a function, instead of eliminating each amplitude $C_{l,k}$ separately.

A similar strategy is employed to compute the $r$ dependence of the amplitude of the other Fourier harmonics $c_l(r,t)$ for $l\not\in\{1,2\}$ as a function of $C_1$ and $C_2$, as also explained below.
This effectively shows that the whole concentration field $c(r,\phi,t)$ can be reexpressed as a perturbation expansion in powers of $C_1$ and $C_2$ after an initial transient relaxation.
We present below a procedure to compute the coefficients of this expansion as explicit functions of $r$ and $\phi$.

\subsection{Perturbation expansion}
\label{scale} 
The following derivation is based on the assumption of $C_1$ and $C_2$ being small.
It is usual for a pitchfork bifurcation to show the perturbation amplitude scaling as $(Pe-Pe_1)^{1/2}$.
Here, however, we have two coupled harmonics that are close to instability, which results in scaling $C_1=O(\epsilon)$ and $C_2=O(\epsilon)$.
The consistency of the evolution equations requires us to admit the following assumptions:
$|Pe-Pe_1|=O(\epsilon)$, $|Pe-Pe_2|=O(\epsilon)$, $\partial_t C_1=O(\epsilon^2)$, $\partial_t C_2=O(\epsilon^2)$, $\delta c_1(r,t)=O(\epsilon^2)$, $\delta c_2(r,t)=O(\epsilon^2)$, $\partial_t\delta c_1(r,t)=O(\epsilon^3)$, $\partial_t\delta c_2(r,t)=O(\epsilon^3)$, $\delta c_0(r,t)=O(\epsilon^2)$, $\partial_t \delta c_0(r,t)=O(\epsilon^3)$.
For $l>2$, we have $\delta c_{l}(r,t)=O(\epsilon^{\lceil l/2\rceil})$ and  $\partial_t \delta c_{l}(r,t)=O(\epsilon^{\lceil l/2\rceil+1})$. The smallness of $\partial_t$ is due to the critical slowing down at the bifurcation point.

\subsection{Adiabatic elimination}

The problem requires us to reduce the partial derivative equations for functions $c_l(r)$ to ordinary differential equations of two scalar amplitudes $C_1$ and $C_2$.
This is performed by applying adiabatic elimination, as explained in this Section.

First, we introduce the right proper function $g_l$ of the operator $\hat L_l(Pe_l)$ such that the corresponding eigenvalue is zero:
\begin{equation}
\label{adj}
\hat L_l^+(Pe_l) g_l(r)=0,
\end{equation}
where $\hat L_l(Pe_l)^+$ is the adjoint operator.
The adjoint operator is defined with respect to the inner product
\begin{equation}
\label{inner}
\langle f,g\rangle=\int\limits_1^R f(r)g(r)^* rdr,
\end{equation}
which is chosen to maintain the self-adjoint property of the diffusion operators $\hat D_l$ subject to the boundary conditions of the functions $c_l(r)$.
The projection condition then reduces to 
\begin{equation}
\label{orthogonality}
\langle g_l,\delta c_l\rangle=0.
\end{equation}

The concentration evolution is split as
\begin{equation}
\label{cdotsplit}
\partial_t c_l(r,t)=\partial_t C_l f_{l}(r)+\partial_t \delta c_l(r,t)\textrm{ for }l\in\{1,2\}
\end{equation}
by the definition (\ref{split}).
We use $g_l$ to isolate the $C_l$ expression from $\partial_t c_l(r,t)$:
\begin{equation}
\label{Cldot}
\partial_t C_l(t)=\frac{\langle g_l,\partial_t c_l(r,t)\rangle}{\langle g_l,f_l\rangle},
\end{equation}
where $\partial_t c(r,t)$ is computed according to eq. (\ref{cldot}).

Equation (\ref{cldot}) for $l\in\{1,2\}$ can be rewritten as
\begin{equation}
\label{cldot2}
\begin{split}
\partial_t c_l(r,t)=&\hat L_l(Pe_l)c_l(r,t)+(Pe-Pe_l)u_l(r)c_l(1)-[\boldsymbol\nabla\boldsymbol\cdot(c(\boldsymbol r)\boldsymbol u(\boldsymbol r))]_l-Pe u_l(r)c_l(1)\\
=&\hat L_l(Pe_l)\underbrace{\delta c_l(r,t)}_{O(\epsilon^2)}+\underbrace{(Pe-Pe_l)}_{O(\epsilon)}u_l(r)\underbrace{c_l(1)}_{O(\epsilon)} \underbrace{-[\boldsymbol\nabla\boldsymbol\cdot(c(\boldsymbol r)\boldsymbol u(\boldsymbol r))]_l-Pe u_l(r)c_l(1)}_{O(\epsilon^2)},
\end{split}
\end{equation}
where the last term is $O(\epsilon^2)$ because $Pe u_l(r)c_l(1)$ represents the $O(\epsilon)$ part of $[\boldsymbol\nabla\boldsymbol\cdot(c(\boldsymbol r)\boldsymbol u(\boldsymbol r))]_l$, as defined in the 
linearization procedure.
Combining eqs. (\ref{cdotsplit}) and (\ref{cldot2}), we obtain
\begin{equation}
\label{Cldot2}
f_l(r)\partial_t C_l(t)=\hat L_l(Pe_l)\delta c_l(r,t)+q_l(r,t)\textrm{ for }l\in\{1,2\},
\end{equation}
where
\begin{equation}
\label{ql}
q_l(r,t)=(Pe-Pe_l)u_l(r)c_l(1)-[\boldsymbol\nabla\boldsymbol\cdot(c(\boldsymbol r)\boldsymbol u(\boldsymbol r))]_l-Pe u_l(r)c_l(1)-\partial_t \delta c_l(r,t)\textrm{ for }l\in\{1,2\}.
\end{equation}

Equation (\ref{Cldot2}) is the main equation of the derivation, which we have solved together with the orthogonality condition (\ref{orthogonality}) to express both $\partial_t C_l$ and $\delta c(r,t)$ as a function of $q_l$.
The definition of $q_l$ in (\ref{ql}) includes the time derivative of $\delta c_l(r)$.
This is not a problem for the subsequent derivation because $\partial_t\delta c_l(r)=O(\epsilon^3)$ and therefore this contribution can be neglected to compute the $O(\epsilon^2)$ terms of $\delta c_l(r,t)$.
The present work only uses the $O(\epsilon^2)$ terms of $\delta c_l(r,t)$ but the proposed method is general enough to compute the higher-order terms as well.
Indeed, once the $O(\epsilon^2)$ terms of $\delta c_l(r,t)$ are obtained as a function of $C_1$ and $C_2$ with coefficients that depend on $r$, the dependence of $\delta c_l(r,t)$ on time is contained only in $C_1(t)$ and $C_2(t)$.
It is thus possible to express the $O(\epsilon^3)$ terms in $\partial_t\delta c_l(r)$ as a function of $C_1$, $C_2$, and their time derivatives.
Those time derivatives can be further reexpressed through $C_1$ and $C_2$ by using the final evolution equations.

The remaining step is to solve the equation (\ref{Cldot2}).
The expression of $\partial_t C_l$ is given by eq. (\ref{Cldot}).
The main challenge is to extract the quasi-static amplitudes $\delta c_l(r,t)$.
Since we have no explicit proper functions for $\hat L_l(Pe_l)$, finding its inverse is far from trivial.
Here the situation is even more complicated because the operator $\hat L_l(Pe_l)$ is not invertible in the first place.

Luckily, here the operator $\hat L_l(Pe_l)$ can be written as
\begin{equation}
\label{niceform}
\hat L_l(Pe_l)=\hat D_l+\boldsymbol u_l\otimes \boldsymbol v_l,
\end{equation}
where $\boldsymbol v_l$ is a distribution defined by the relation $\langle \boldsymbol f,\boldsymbol v_l\rangle=Pe_lf(1)$ (aka Dirac delta function).
It is classically known that it is possible to invert explicitly the operators of form (\ref{niceform}), provided the inverse operator for $D_l$ is known:
\begin{equation}
\label{inverse}
(\hat D_l+\boldsymbol u_l\otimes \boldsymbol v_l)^{-1}=\hat D_l^{-1}-\frac{(\hat D_l^{-1}\boldsymbol u_l)\otimes(\boldsymbol v_l\hat D_l^{-1})}{1+\langle v_l,\hat D_l^{-1}\boldsymbol u_l\rangle}
\end{equation}

This is the case for our study as the inverse of the diffusion operator can be represented in an integral form with a simple kernel, as shown in Appendix \ref{app::diffusion}.
The remaining difficulty is the lack of a well-defined inverse for the operator $\hat L_l(Pe_l)$, which we overcome, as explained in Appendix \ref{app::advection-diffusion}.
The resulting solution of eq. (\ref{Cldot2}) reads:
\begin{equation}
\label{deltac12}
\delta c_l=-\hat D_l^{-1}q_l+\frac{\langle v_l,\hat D_l^{-1}q_l\rangle}{\langle v_l,\hat D_l^{-1}f_l\rangle}\hat D_l^{-1}f_l+p_lf_l,
\end{equation}
where the coefficient $p_l$ is chosen to satisfy eq. (\ref{orthogonality}):
\begin{equation}
\label{psolution}
p_l=\frac{\langle v_l,D^{-2}q_l\rangle\langle v_l,D_l^{-1}f_l\rangle-\langle v_l,D_l^{-1}q_l\rangle\langle v_l,D_l^{-2}f_l\rangle}{\langle v_l,D_l^{-1}f_l\rangle^2}
\end{equation}
Note that we have used here that $f_l=D_l^{-1}u_l$ and $g_l=D^{-1}v_l$, as shown in the Appendix \ref{app::advection-diffusion}.

The functions $\delta c_l(r,t)$ for $l\in\{1,2\}$ satisfy equation (\ref{cldot2}), which can be solved according to eqs. (\ref{deltac12}) and (\ref{psolution}).
The remaining functions $\delta c_l(r,t)$ for $l\not\in\{1,2\}$ satisfy similar equations
\begin{equation}
\label{Cldot21}
0=\hat L_l(Pe)\delta c_l(r,t)+q_l(r,t)\textrm{ for }l\not\in\{1,2\},
\end{equation}
where the function $q_l$ is given by
\begin{equation}
\label{ql2}
q_l(r,t)=-[\boldsymbol\nabla\boldsymbol\cdot(c(\boldsymbol r)\boldsymbol u(\boldsymbol r))]_l-Pe u_l(r)c_l(1)-\partial_t \delta c_l(r,t)\textrm{ for }l\not\in\{1,2\}.
\end{equation}
Since the kernel $L_l(Pe_l)$ is not singular for $l\not\in\{1,2\}$ and $Pe$ close to $Pe_1$ and $Pe_2$, eq. (\ref{inverse}) is used to compute the functions $\delta c_l(r,t)$ as the solutions of eq. (\ref{Cldot21}) for all $l\not\in\{1,2\}$.

This completes the proof that the full partial differential equations governing the evolution of the concentration field $c(r,t)$ can be reduced to a system of two differential equations for $C_1$ and $C_2$.
The practical implementation of the derivation procedure and the explicit expressions for some intermediate results and the final equations are given in the next Section.

 \section{The derivation procedure} 

We use an iterative procedure obtaining the expansions of $\partial_t C_1$, $\partial_t C_2$, and $\delta c_l$ one order of $\epsilon$ at a time.
The first step is to neglect the $O(\epsilon^2)$ terms in $c(r,\phi)$, which allows us to compute the $O(\epsilon^2)$ terms in $\partial_t C_1$, $\partial_t C_2$, and $\delta c_l$.
We then substitute the obtained expressions of $\delta c_l$ into $c(r,\phi)$, keeping the terms up to $O(\epsilon^2)$ order this time, which is sufficient to obtain the order of $O(\epsilon^3)$ terms in $\partial_t C_1$, $\partial_t C_2$, and $\delta c_l$.
This procedure is continued ad infinitum but we stop at computing the $O(\epsilon^3)$ terms of $\partial_t C_1$ and $\partial_t C_2$, which is sufficient for our purposes.

According to the eq. (\ref{split}) and the order of magnitude analysis (see section \ref{scale}), the concentration field can be written as
\begin{equation}
\label{clong}
\begin{split}
c(r,\phi,t)=&c_0(r)+\left[\left(C_1 e^{i\phi}+C_1^*e^{-i\phi}\right)f_1(r)+\left(C_2 e^{2i\phi}+C_2^*e^{-2i\phi}\right)f_2(r)\right]+\\
&\left[\delta c_0(r)+\sum\limits_{l=1}^4 \left(\delta c_l(r)e^{il\phi}+\delta c_l(r)^*e^{-il\phi}\right)\right]+O(\epsilon^3)
\end{split}
\end{equation}
where $C_1$ and $C_2$ are time-dependent complex amplitudes of the two modes that are close to instability and $\delta c_k(r)$ are functions of $r$ with coefficients that depend on $C_1$, $C_2$, $R$, and $Pe$.

Each iteration consists in the following steps:
\begin{enumerate}
\item The concentration field (\ref{clong}) defines the stream function according to eq. (\ref{psil}), from which the fluid velocity field is calculated.
\item Knowing the velocity field and the concentration, the advection term $\boldsymbol\nabla\boldsymbol\cdot(c(\boldsymbol r)\boldsymbol u(\boldsymbol r))$ is computed.
\item The advection term is then decomposed into Fourier harmonics which yields the amplitudes $[\boldsymbol\nabla\boldsymbol\cdot(c(\boldsymbol r)\boldsymbol u(\boldsymbol r))]_l$.
\item These amplitudes are then used to compute the $q_l$ terms according to eqs. (\ref{ql}) and (\ref{ql2}).
\item $\partial_t C_1$ and $\partial_t C_2$ are computed at a given order according to eqs. (\ref{Cldot}).
\item The quasi-static values of $\delta c_l$ are computed according to eqs. (\ref{deltac12}) and (\ref{psolution}) for $l\in\{1,2\}$ and according to eq. (\ref{inverse}) for $l\in\{1,2\}$.
\end{enumerate}
The last step can be omitted for the final iteration.

\section{Evolution equations for $C_1(t)$ and $C_2(t)$ to leading order}
We first start with the derivation to order $\epsilon^2$. The starting point is to use (\ref{Cldot2}). Multiplying this equation by $g_l$ and integrating both sides according to the scalar product (\ref{inner})
we obtain
\begin{equation}
\label{Cldot2p}
\partial_t C_l(t)={\langle  g_l, q_l  \rangle  \over \langle g_l,f_l\rangle  } ,
\end{equation}
where we have used  $\langle  g_l, \hat L_l\delta c_l  \rangle=\langle \hat L_l^+ g_l(r),\delta c_l\rangle =0$, by virtue of (\ref{adj}). Since $\partial_t \delta c_l(r,t)=O(\epsilon^3)$, this term in $q_l$ (Eq. (\ref{ql})) does not enter to order $\epsilon^2$. The next step is to insert  (\ref{clong}) into $q_l$ (Eq. (\ref{ql})) and report the resulting expression into (\ref{Cldot2p}). Note that $\delta c_l$ in (\ref{clong}) does not enter to this order neither since it produces $O(\epsilon^3)$ contribution to (\ref{Cldot2p}) (see also (\ref{Cldot2}) for orders in $\epsilon$). We are thus left, on the right hand side of (\ref{Cldot2p}), with scalar products involving $f_l$ and $g_l$ (and $u_l$, which is a known function, see (\ref{ul}) ) with prefactors containing linear and quadratic terms of $C_l(t)'s$. The inner product calculation requires, in principle, the knowledge of $f_l$ and $g_l$. 

The linear stability of the solution (\ref{c0}) is governed by the eigenvalues of the linear operators $\hat L_l$ in eq. (\ref{linear}).
The critical Peclet numbers are
\begin{equation}
\label{Pe_1}
Pe_1=- \frac{2 \left(R^{2} + 1\right)}{R^{2} - \left(R^{2} + 1\right) \ln{\left(R \right)} - 1}
\end{equation}
\begin{equation}
\label{Pe_2}
Pe_2=- \frac{R^{4} + 1}{- \frac{R^{4}}{4} + R^{2} - \ln{\left(R \right)} - \frac{3}{4}}
\end{equation}
and the corresponding proper functions are
\begin{equation}
\label{f1}
f_1(r)=\frac{R^{2}-r^{2}}{2r \left(R^{2} + 1\right)} + \frac{\left(r^2 + 1\right) \ln{\left(\frac{r}{R} \right)}}{4r}
\end{equation}
\begin{equation}
\label{f2}
f_2(r)=\frac{\left(- R^{2} + r^{2}\right) \left(2 R^{2} - 2 \left(\frac{R^{2}}{r^{2}} + 1\right) \ln{\left(R \right)} - 1 - \frac{R^{2} + 2}{r^{2}}\right)}{4 R^{4} + 4} + \frac{\ln{\left(\frac{r}{R} \right)}}{2 r^{2}}.
\end{equation}
Here we have computed $f_l$ as $D_l^{-1}u_l$ and $Pe_l$ is obtained from condition $\langle v_l,D^{-1} u_l\rangle=1$, as shown in Appendix \ref{app::advection-diffusion}. It turns out we do not need the explicit expression of $g_l$.
Indeed, noting that because  $g_l=D^{-1}v_l$ (see  Appendix \ref{app::advection-diffusion}), any inner product in the form $\langle g_l, H(r) \rangle$ can be written as 
\begin{equation}
\label{dist}
\langle g_l, H(r) \rangle= \langle \hat D_l^{-1}v_l, H(r)\rangle =\langle v_l,\hat D_l^{-1} H(r)\rangle=Pe_l [ \hat D_l^{-1} H(r)]_{r=1}  
\end{equation} 
Recall that $\hat D_l$ is self adjoint and so is  its inverse.  The calculation of $\hat D_l^{-1}$  is performed in 
Appendix \ref{app::diffusion}. We are now in a position to calculate the inner product in (\ref{Cldot2p}). Consider the case $l=1$, and collect the linear term in $c_1$ in (\ref{ql}), which is given by $(Pe-Pe_1) u_1 c_1(1)=(Pe-Pe_1)C_1(t) u_1(r) f_1(1)$ (where $u_1(r)$ is given by given by (\ref{ul})). Using (\ref{Cldot2p}), we obtain from the right hand side  
\begin{equation} 
(Pe-Pe_1)C_1(t)f_1(1) {\langle g_1(r) , u_1(r)\rangle \over \langle g_1(r) , f_1(r)\rangle} =(Pe-Pe_1)C_1(t)f_1(1)  {[ \hat D_1^{-1} u(r)]_{r=1} \over [ \hat D_1^{-1} f_1(r)]_{r=1}  } \end{equation} 
where we have used (\ref{dist}).  In Appendix \ref{app::diffusion} wee show how to calculate $D_1^{-1}$, and the above expression can easily be evaluated as a function of $R$.  Reporting this into 
(\ref{Cldot2p}) yields the $C_1(t)$ equation to linear  order. The next term from $q_l$  is quadratic in $c_l$ and comes from the combination of second (advection term) and third terms in (\ref{ql}). It reads, after using the expressions of $u_r$ and $u_\theta$ (see (\ref{stream}), where $\psi_l$  is given by (\ref{psil}))
\begin{equation} 
\sum _{m\ne 1}  {m c_m(1,t) Pe \over 2 r^{\vert m\vert +1} }  \left[ (r^2-1)m \frac{\partial c_{1 -m}( r,t) }{\partial r}  + (m-1)  (2r^2 +(1-r^2) \vert m\vert ) c _{1 -m} (r,t) \right]
\end{equation}
By using $c_l(r,t)=C_l(t) f_l(r)$, and retaining only first and second harmonics, it is easy to see that the result reads as $C_1^* C_2 h(r)$, where $h(r)$ is a function of $r$ only (it is a combination of $f_1$ and $f_2$ and their derivatives with respect to $r$). Once this expression is injected into (\ref{Cldot2p}) on the right hand side we obtain 

\begin{equation} 
C_1(t)^* C_2(t) {\langle g_1(r) , h(r)\rangle \over \langle g_1(r) , f_1(r)\rangle} = C_1(t)^* C_2(t)  {[ \hat D_1^{-1} h(r)]_{r=1} \over [ \hat D_1^{-1} f_1(r)]_{r=1}  } \end{equation} 
The last term can  easily be obtained as algebraic rational functions of $R$ and $\ln R$ (see Appendix \ref{app::advection-diffusion}). The same reasoning can be made for the equation of $C_2$. The nonlinear term is found to be proportional to $C_1(t)^2$.
Collecting linear and nonlinear terms in $C_1$ and $C_2$, the resulting system of equations  to the second order is found to be given by 
\begin{subequations}
\label{Cdot}
\begin{align}
\dot C_1&=\sigma_1 C_1+\alpha_1 C_2C_1^*+O(\epsilon^3),\\
\dot C_2&=\sigma_2 C_2+\alpha_2 C_1^2+O(\epsilon^3),
\end{align}
\end{subequations}
where different coefficients are functions of $R$ and are listed in Appendix \ref{coeff}.

\section{Equations for $C_1$ and $C_2$ to next order}
The next order terms turn out to be essential for nonlinear saturation. We need thus to extend the derivation to $O(\epsilon^3)$.
 We use (\ref{clong}) and insert it into (\ref{Cldot2p}). Taking into account $\delta c_l$ in  (\ref{clong}) will lead to higher order terms. 
 $\delta c_l$ is given by (\ref{deltac12}) and (\ref{psolution}) for $l\in\{1,2\}$ and by (\ref{inverse}) for $l\not\in\{1,2\}$.
  Inserting these solutions into (\ref{clong}) allows us, by using (\ref{Cldot2p}), to obtain the desired terms. 
The cubic terms are of the form  $\lvert C_2\rvert ^2 C_1$ for the equation of $C_1$   and $\lvert C_1\rvert^2 C_2$ for the equation of $C_2$. There is also a  contribution of $\delta c_l$   in the form of  $\partial _t \delta c_l $ in definition of $q_l$. Since the time derivative is small (critical slowing down), the quadratic contribution arising from $\delta c_l$ are sufficient. For example the equation for $C_1$ yields terms in the form $\partial _t (C_1^* C_2)= C_1^* \partial_t C_2+ C_2 \partial_t C_1^*$. Using (\ref{Cdot}) we can express these terms as quadratic and cubic terms.  The final set of equation takes the form
\begin{subequations}
\label{Cdot3}
\begin{align}
\dot C_1&=\sigma_1 C_1+\alpha_1 C_2C_1^*- \beta_1 | C_2|  ^2 C_1 -\gamma_1 | C_1|  ^2 C_1 ,\\
\dot C_2&=\sigma_2 C_2+\alpha_2 C_1^2-\beta_2  | C_1|  ^2 C_2  -  \gamma_2 | C_2|  ^2 C_2 ,
\end{align}
\end{subequations}
The   coefficients of the cubic terms  have  expressions in $R$ which become too involved.
For this reason, we have resorted to computing them for a given numerical value of $R$, becoming just real numbers. In Appendix \ref{coeff} we plot these coefficients as functions of $R$.

\section{Comparison between the full model and the reduced one}
\begin{figure}
\begin{center}
\includegraphics[width=0.9\linewidth]{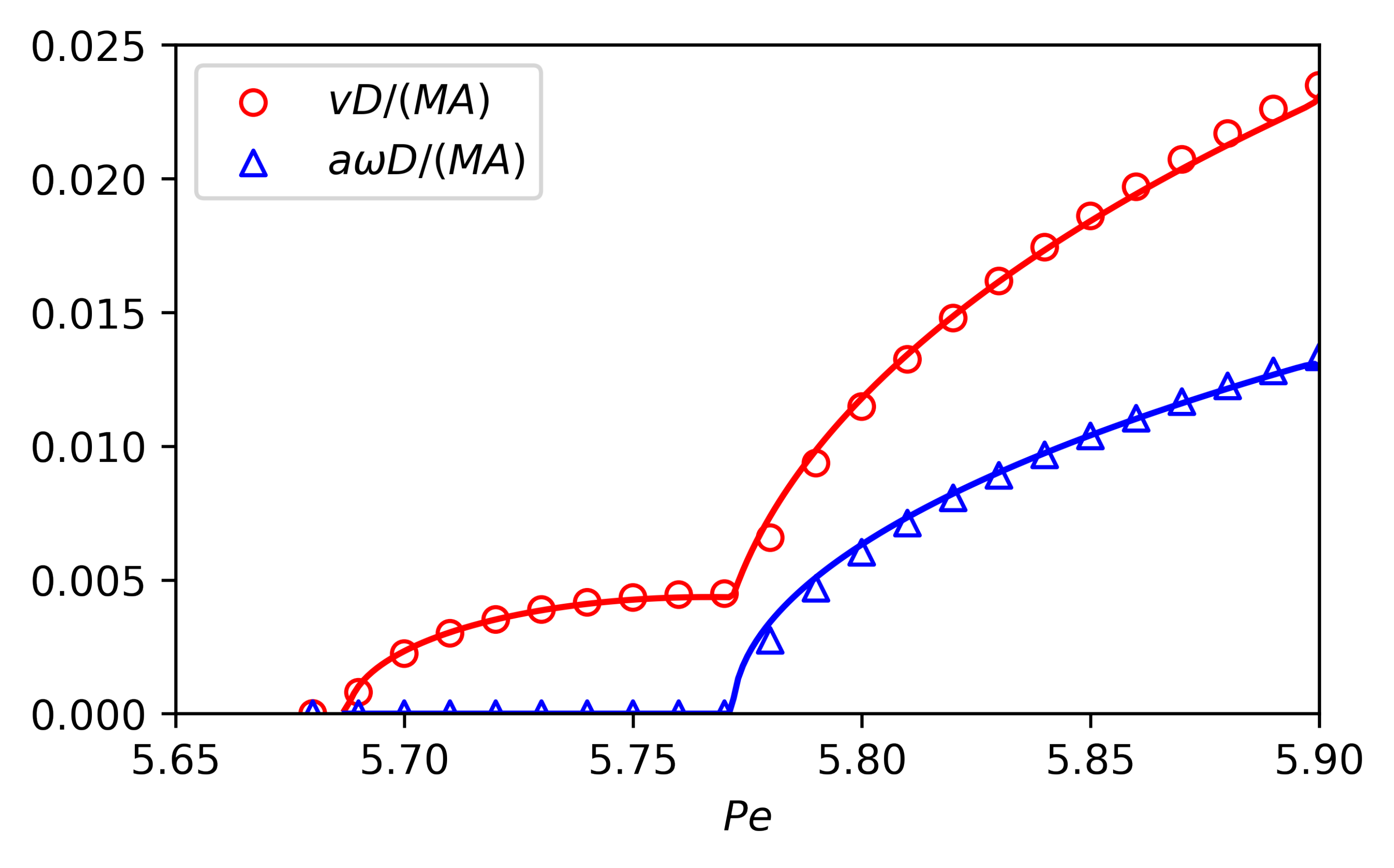}
\end{center}
\caption{\label{compare}
Comparison between the analytical prediction and the direct numerical simulation of the particle velocity and angular velocity in stationary, straight, and circular phases.
The system size is set to $R=3.25$.
The solid curves are obtained by a direct numerical solution of the $C_1$, $C_2$ equations.
The symbols are the results of the full numerical simulations.
In both cases the steady-state values are obtained by running the simulations for a long time until (on the order of $10^5a^2/D$).
The horizontal axis is cut at $Pe=5.9$ above which the absolute value of the velocity does not seem to reach a steady state.
}
\end{figure}

Here we would like to confront the reduced model (\ref{Cdot3}) with the full model (as described in Section \ref{fullmodel}).
The full model is solved using finite difference discretization.
The reduced model (\ref{Cdot3}) is a set of ordinary differential equations, the numerical solution of which is straightforward.  
The results of the comparison are shown in Fig. \ref{compare} for the steady-state values of the swimming speed and the angular velocity of the particle (a nonzero value of the angular velocity corresponds to a circular trajectory).
As can be seen, the reduced $C_1$ and $C_2$ model is in quantitative agreement with the full numerical simulation.

\begin{figure}
\begin{center}
\includegraphics[width=0.9\linewidth]{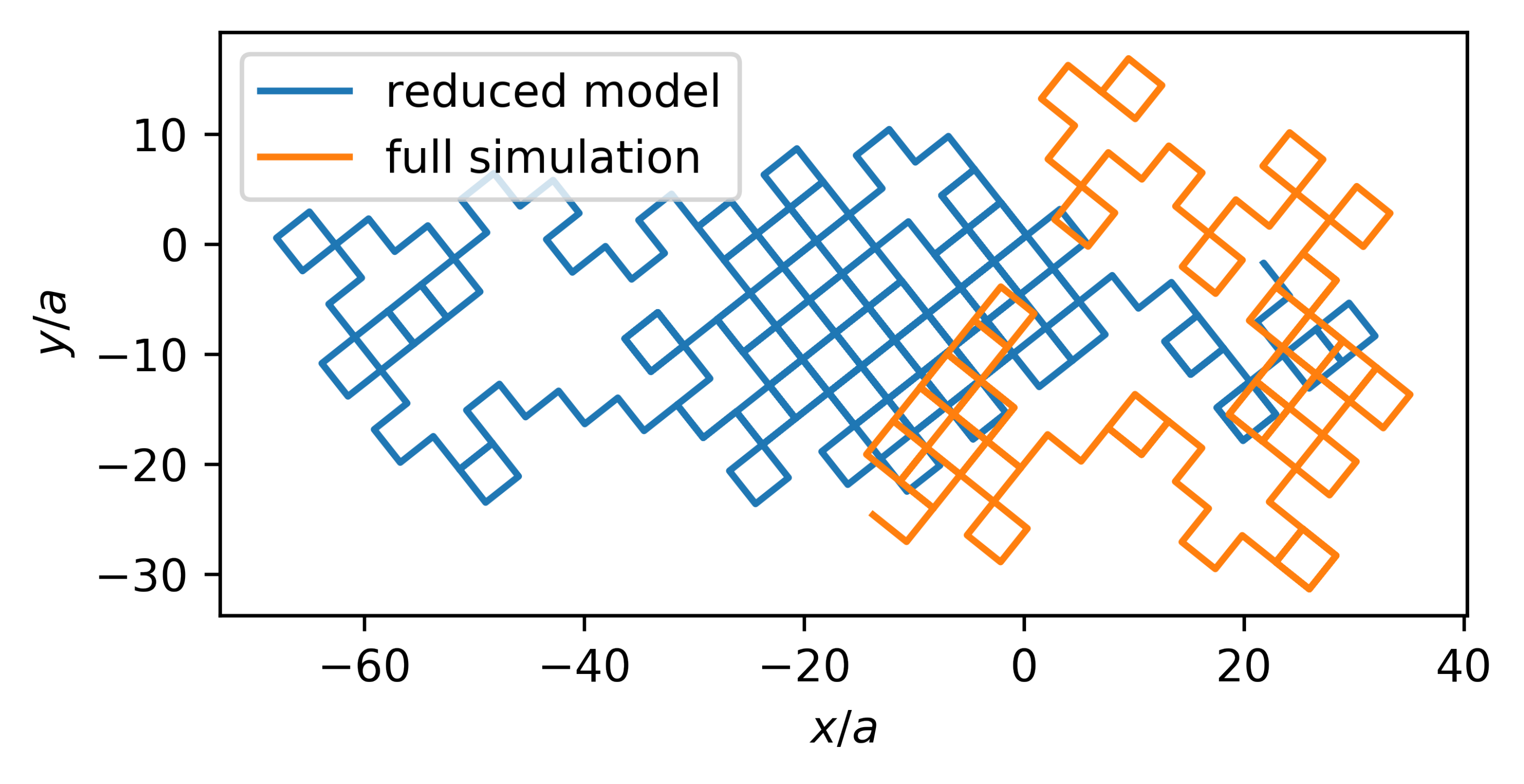}
\end{center}
\caption{\label{chaos}
An irregular trajectory obtained from solution of (\ref{Cdot3}) for $Pe=5.95$.
}
\end{figure}

The solution of the $C_1$ and $C_2$ model shows that as the P\'eclet number is increased, the particle shows non-motile solution as the stable fixed point for $Pe<Pe_1$, straight motion (velocity is constant in time and finite, but angular velocity is equal to zero) for $Pe^*>Pe>Pe_1$, and a circular trajectory (both the swimming speed and the angular velocity remain fixed in time) for $Pe^{**}>Pe>Pe^*$.
No stable fixed point solution could be obtained for $Pe>Pe^{**}$. Here we have $Pe^*\approx 5.77$ and $Pe^{**}\approx 5.9$. By increasing further $Pe$ we find that the solution becomes irregular, pointing to the occurrence of chaos (Fig.\ref{chaos} .
This succession of transitions from one trajectory type to another reproduces the prediction of the phenomenological theory. The circular trajectory and chaotic one have been also obtained by numerical simulation in \cite{Hu2019} using the full model. In a recent work \cite{MisbahC1C2}  by taking the set of equations (\ref{Cdot3}) as a phenomenological model (extracted on the basis of symmetries,   without reference to any given basic model)   it has been shown that it reproduces a variety of solutions going from  straight, circular to chaotic trajectoriees. The circular trajectory could even be obtained analytically from (\ref{Cdot3}).

\section{Conclusion}
We have extracted from  a phoretic model in 2D, including advection and diffusion,  a reduced model in terms of ordinary differential  equations for the first two Fourier modes. This model captures, in a much less numerically expensive manner,  essential features encountered in the full model. The derivation has been performed for the two first modes to cubic order. Extension to a higher number of modes can also be performed, as well as going to higher orders than cubic terms. This will be essential if one wishes to have a wider applicability of the method. The derivation  can be extended to 3D without any additional conceptual complications. The method has been exemplified for a specific phoretic model. However, the same technique can be used for any swimmer powered by one or many chemical fields. Examples of great topicality are found in models of cell motility, involving acto-myosin kinetics inside the cell, outside and on the cell  membrane, cortex flow as well as flow in the suspending and internal  fluids \cite{hawkins2011spontaneous,Voituriez2016,Farutin2019}. The present theory constitutes a precious alternative to full numerical simulations. In addition, as the reduced  model equations have a form which depends only on symmetries, different active entities can be   put together in a universal framework, where different models can be described by  the same reduced evolution equations (\ref{Cdot3}), and their differences will be encoded in the values of the numerical coefficients only.

\section{Acknowledgments}

We thank CNES (Centre National d'Etudes Spatiales) for financial support and for having access to microgravity data, and the French-German university program "Living Fluids" (grant CFDA-Q1-14) for financial support.\appendix

\section{\label{app::diffusion}Inverting the diffusion operators}

A solution to the equation
\begin{equation}
\label{diffusion}
\frac{1}{r}\frac{d}{dr}\left(r\frac{dF}{dr}\right)-l^2F=G(r)
\end{equation}
can be written as
\begin{equation}
\label{kernel}
F(r)=
\begin{cases}
\frac{1}{2l}\left[r^{-l}\int^r G(\rho)\rho^{1+l}d\rho-r^l\int^r G(\rho)\rho^{1-l}d\rho\right]+A_1 r^l+A_2r^{-l}&\textrm{ for }l>0\\
\ln r\int^r G(\rho)\rho d\rho-\int^r \rho\ln \rho G(\rho)d\rho+A_1+A_2\ln r&\textrm{ for }l=0
\end{cases},
\end{equation}
where the constants $A_1$ and $A_2$ are calculated by imposing the boundary conditions
\begin{equation}
\label{boundary}
\partial_rF(1)=0\,\,\,F(R)=0.
\end{equation}
The problem considered here requires solving eq. (\ref{diffusion}) with functions $G(r)$ written as a combination of integer powers of $r$ and $\ln r$.
This allows us to calculate the integrals in (\ref{kernel}) analytically.

\section{\label{app::advection-diffusion}Inverting the advection-diffusion operator}

Suppose we have a system of equations
\begin{equation}
\label{matheq}
\begin{split}
&\mu \boldsymbol f=(M+\boldsymbol u\otimes \boldsymbol v).\boldsymbol x +\boldsymbol y\\
&\langle \boldsymbol g,\boldsymbol x\rangle =0,
\end{split}
\end{equation}
where $\mu$ is a number, $M$ is a matrix, $\boldsymbol u,$ $\boldsymbol v$, $\boldsymbol x$, $\boldsymbol y$, $\boldsymbol f$, $\boldsymbol g$ are vectors, and
\begin{equation}
\label{matheigen}
\begin{split}
&(M+\boldsymbol u\otimes \boldsymbol v).\boldsymbol f=0,\\
&\boldsymbol g.(M+\boldsymbol u\otimes \boldsymbol v)=0.
\end{split}
\end{equation}
The first equation (\ref{matheq}) corresponds to eq.(\ref{Cldot2})  where $\mu$ represents $\partial _t C_l$, $M+\boldsymbol u\otimes \boldsymbol v$ is the linear operator $\hat L (Pe_l)$ (see eq; \ref{niceform}), ${ \boldsymbol x}$ represents $\delta c_l$ and ${\boldsymbol y}$ represents $q_l$. The second equation represents the orthogonality condition (\ref{orthogonality}). Equation (\ref{matheigen}) represents the eigenvalue problem  $\hat L (Pe_l) f_l (r)=0$ and its adjoint (\ref{adj}). 

The goal of this Section is to find the solution of the system (\ref{matheq}) representing $\boldsymbol x$ and $\mu$ as a function of $M^{-1}$, $\boldsymbol u$, $\boldsymbol v$, and $\boldsymbol y$.
First, we note the following relations:
\begin{equation}
\label{fgdef}
\boldsymbol f\propto M^{-1}.\boldsymbol u,\,\,\,\boldsymbol g\propto \boldsymbol v.M^{-1},\,\,\,\langle \boldsymbol v,M^{-1}.\boldsymbol u\rangle=-1,
\end{equation}
which we obtain from eqs. (\ref{matheigen}).
Multiplying the first equation in (\ref{matheq}) by $M^{-1}$, we get
\begin{equation}
\label{matheq2}
\mu M^{-2}.\boldsymbol u=[I+(M^{-1}.\boldsymbol u)\otimes \boldsymbol v].\boldsymbol x +M^{-1}.\boldsymbol y,\\
\end{equation}
whence
\begin{equation}
\label{matheq3}
\boldsymbol x=\mu M^{-2}.\boldsymbol u-(M^{-1}.\boldsymbol u)\langle \boldsymbol v,\boldsymbol x\rangle -M^{-1}.\boldsymbol y,\\
\end{equation}
Equation (\ref{matheq3}) implies the following ansatz for $\boldsymbol x$:
\begin{equation}
\label{ansatz}
\boldsymbol x=-M^{-1}.\boldsymbol y+pM^{-1}.\boldsymbol u+qM^{-2}.\boldsymbol u,\\
\end{equation}
where $p$ and $q$ are two numbers to be determined.
It is then straightforward to substitute eq. (\ref{ansatz}) into eq. (\ref{matheq3}) to get the values of $\mu$ and $q$ as:
\begin{equation}
\label{qmusol}
q=\mu=\frac{\langle \boldsymbol v,M^{-1}.\boldsymbol y\rangle}{\langle \boldsymbol v,M^{-2}.\boldsymbol u\rangle}
\end{equation}
The value of $p$ is determined from the second equation of (\ref{matheq}).
It is convenient that the solution (\ref{qmusol}) use the vector $\boldsymbol v$ only as part of $\langle \boldsymbol v,\boldsymbol\cdot\rangle$ which allows us to use its definition $\langle \boldsymbol f,\boldsymbol v\rangle=Pef(1)$.
\section{\label{coeff} Coefficients of the reduced model}
Here we list the coefficients of the linear and quadratic terms and plot those of the cubic term.

\begin{equation}
\label{sigma1}
\sigma_1=\frac{16 (Pe-Pe_{1}) \left(R^{2} \ln{\left(R \right)} - R^{2} + \ln{\left(R \right)} + 1\right)^{2}}  {3 R^{6} + 8 R^{4} \ln{\left(R \right)}^{2} - 40 R^{4} \ln{\left(R \right)} + 19 R^{4} + 8 R^{2} \ln{\left(R \right)}^{2} + 4 R^{2} \ln{\left(R \right)} - 35 R^{2} + 12 \ln{\left(R \right)} + 13}.
\end{equation}
\begin{equation}
\label{alpha1}
\alpha_1=- \frac{\begin{matrix}2 \left(R^{2} + 1\right) \left(20 R^{10} \ln{\left(R \right)}^{2} - 40 R^{10} \ln{\left(R \right)} + 21 R^{10} + 8 R^{8} \ln{\left(R \right)}^{2} + 48 R^{8} \ln{\left(R \right)} - 63 R^{8} \right. &\\ \left. + 16 R^{6} \ln{\left(R \right)}^{3} - 48 R^{6} \ln{\left(R \right)}^{2} + 56 R^{6} \ln{\left(R \right)} + 46 R^{6} + 32 R^{4} \ln{\left(R \right)}^{3} - 56 R^{4}  \ln{\left(R \right)}^{2} \right. &\\ \left. - 8 R^{4} \ln{\left(R \right)} + 6 R^{4} + 16 R^{2} \ln{\left(R \right)}^{3} - 20 R^{2} \ln{\left(R \right)}^{2} - 48 R^{2} \ln{\left(R \right)} - 3 R^{2} - 8 \ln{\left(R \right)} - 7\right)\end{matrix}}
{ \begin{matrix} R^{2} 
\left(R^{4} + 1\right) \left(R^{2} \ln{\left(R \right)} - R^{2} + \ln{\left(R \right)} + 1\right) \left(3 R^{6} + 8 R^{4} \ln{\left(R \right)}^{2} - 40 R^{4} \right. &\\ \left. \ln{\left(R \right)} + 19 R^{4} + 8 R^{2} \ln{\left(R \right)}^{2} + 4 R^{2} \ln{\left(R \right)} - 35 R^{2} + 12 \ln{\left(R \right)} + 13\right) \end{matrix}}
\end{equation}

\begin{equation}
\label{sigma2}
\sigma_2=\frac{3 (Pe-Pe_{2}) \left(R^{4} - 4 R^{2} + 4 \ln{\left(R \right)} + 3\right)^{2}}{12 R^{8} \ln{\left(R \right)} - 13 R^{8} + 16 R^{6} \ln{\left(R \right)} + 8 R^{6} + 12 R^{4} \ln{\left(R \right)} + 6 R^{4} - 8 R^{2} + 8 \ln{\left(R \right)} + 7}
\end{equation}

\begin{equation}
\label{alpha2}
\alpha_2=\frac{\begin{matrix} 3 \left(R^{4} + 1\right)^{2} \left(R^{2} \ln{\left(R \right)} - R^{2} + \ln{\left(R \right)} + 1\right) \left(6 R^{6} \ln{\left(R \right)} - 7 R^{6} + 6 R^{4} \ln{\left(R \right)} \right.&\\\left. + 9 R^{4} + 6 R^{2} \ln{\left(R \right)} - 9 R^{2} + 6 \ln{\left(R \right)} + 7\right)  \end{matrix}} {\begin{matrix} \left(R^{2} + 1\right)^{2} \left(R^{4} - 4 R^{2} + 4 \ln{\left(R \right)} + 3\right) \left(12 R^{8} \ln{\left(R \right)} \right.&\\\left. - 13 R^{8} + 16 R^{6} \ln{\left(R \right)} + 8 R^{6} + 12 R^{4} \ln{\left(R \right)} + 6 R^{4} - 8 R^{2} + 8 \ln{\left(R \right)} + 7\right) \end{matrix}}
\end{equation} 


\section{Explicit expression}

Here we list the explicit epxressions for $\partial_t C_1$ and $\partial_t C_2$ for $R=3.25$.
We first have simplified the expressions assuming $Pe$ to be close to the critical P\'eclet number $Pe_c=5.9561$, which is the critical P\'eclet number for $R=R_c=3.17493$, such that $Pe_1(R_c)=Pe_2(R_c)=Pe_c$.
We then define $\Delta P=Pe-Pe_c=O(\epsilon)$ and truncate all expressions keeping only terms of order $O(\epsilon^3)$ or higher.

\begin{subequations}
\begin{align}
\label{C1dot_3}
\partial_t C_1&= C_{1} \left(0.0484 + 0.1799 \Delta Pe- 0.00244 \Delta Pe^{2} \right)- C^{*}_{1} C_{2} \left( 1.227+0.2467 \Delta Pe \right)-\\
& 0.2016 C_{1}|C_1|^2 - 3.077 C_{1} |C_2|^2 +O(\epsilon^4)\notag,\\
\label{C2dot_3}
\partial_t C_2&=C_{2} \left(0.0411+ 0.3701 \Delta Pe- 0.00123 \Delta Pe^{2} \right)+C_{1}^{2} \left(0.5921+0.0905 \Delta Pe \right)-\\
 & 3.1214 |C_1|^2 C_{2} - 1.7893 |C_2|^2C_{2} \notag.
\end{align}
\end{subequations}
%
\end{document}